\magnification\magstephalf
\input epsf
\overfullrule 0pt

\font\rfont=cmr9 at 9 true pt
\def\ref#1{$^{\hbox{\rfont {[#1]}}}$}


\font\fourteenbf=cmbx12 scaled\magstep1

\font\tenbfit=cmbxti10
\font\sevenbfit=cmbxti10 at 7pt
\font\fivebfit=cmbxti10 at 5pt
\newfam\bfitfam 
\textfont\bfitfam=\tenbfit  \scriptfont\bfitfam=\sevenbfit
\scriptscriptfont\bfitfam=\fivebfit

\font\eightit=cmti8

\font\tenbfit=cmbxti10
\font\sevenbfit=cmbxti10 at 7pt
\font\fivebfit=cmbxti10 at 5pt
\newfam\bfitfam 
\textfont\bfitfam=\tenbfit  \scriptfont\bfitfam=\sevenbfit
\scriptscriptfont\bfitfam=\fivebfit

\font\tenbit=cmmib10
\newfam\bitfam
\textfont\bitfam=\tenbit%

\font\tenmbf=cmbx10
\font\sevenmbf=cmbx7
\font\fivembf=cmbx5
\newfam\mbffam
\textfont\mbffam=\tenmbf \scriptfont\mbffam=\sevenmbf
\scriptscriptfont\mbffam=\fivembf

\font\tenbsy=cmbsy10
\newfam\bsyfam 
\textfont\bsyfam=\tenbsy%

  
\def\e{\epsilon}  
  \def\la{\lambda}
 \def\La {\Lambda} 
\def\pd {\partial}
\def\pmb#1{\setbox0=\hbox{#1}
 \kern.05em\copy0\kern-\wd0 \kern-.025em\raise.0433em\box0 }
\def\bpi{{\pmb{$ \pi$}}}
\def\bg{{\pmb{$ \gamma$}}}
\def\slash{/\kern-.5em}


\def \quarter {{\textstyle {1 \over 4}}}

 %


\def\boxit#1{\vbox{\hrule\hbox{\vrule\kern1pt\vbox
{\kern1pt#1\kern1pt}\kern1pt\vrule}\hrule}}

\def\h{\hfill\break}
\parskip=6pt
\parindent=0pt
\hsize=17truecm\hoffset=-5truemm
\vsize=22.5truecm
\def\footnoterule{\kern-3pt
\hrule width 17truecm \kern 2.6pt}


\catcode`\@=11 

\def\nolabels{\def\wrlabeL##1{}\def\eqlabeL##1{}\def\reflabeL##1{}}
\def\writelabels{\def\wrlabeL##1{\leavevmode\vadjust{\rlap{\smash%
{\line{{\escapechar=` \hfill\rlap{\sevenrm\hskip.03in\string##1}}}}}}}%
\def\eqlabeL##1{{\escapechar-1\rlap{\sevenrm\hskip.05in\string##1}}}%
\def\reflabeL##1{\noexpand\llap{\noexpand\sevenrm\string\string\string##1}}}
\nolabels
\global\newcount\refno \global\refno=1
\newwrite\rfile
\def\defref{$^{{\hbox{\rfont [\the\refno]}}}$\nref}
\def\nref#1{\xdef#1{\the\refno}\writedef{#1\leftbracket#1}%
\ifnum\refno=1\immediate\openout\rfile=refs.tmp\fi
\global\advance\refno by1\chardef\wfile=\rfile\immediate
\write\rfile{\noexpand\item{#1\ }\reflabeL{#1\hskip.31in}\pctsign}\findarg}
\def\findarg#1#{\begingroup\obeylines\newlinechar=`\^^M\pass@rg}
{\obeylines\gdef\pass@rg#1{\writ@line\relax #1^^M\hbox{}^^M}%
\gdef\writ@line#1^^M{\expandafter\toks0\expandafter{\striprel@x #1}%
\edef\next{\the\toks0}\ifx\next\em@rk\let\next=\endgroup\else\ifx\next\empty%
\else\immediate\write\wfile{\the\toks0}\fi\let\next=\writ@line\fi\next\relax}}
\def\striprel@x#1{} \def\em@rk{\hbox{}} 
\def\lref{\begingroup\obeylines\lr@f}
\def\lr@f#1#2{\gdef#1{\defref#1{#2}}\endgroup\unskip}
\newwrite\lfile
{\escapechar-1\xdef\pctsign{\string\%}\xdef\leftbracket{\string\{}
\xdef\rightbracket{\string\}}}

\def\writestop{\def\writestoppt{\immediate\write\lfile{\string\p
ageno%
\the\pageno\string\startrefs\leftbracket\the\refno\rightbracket%
\string\def\string\secsym\leftbracket\secsym\rightbracket%
\string\secno\the\secno\string\meqno\the\meqno}\immediate\closeout\lfile}}
\def\writestoppt{}\def\writedef#1{}
\catcode`\@=12 
\def\v{{\bf v}}
\def\u{{\bf u}}
\def\U{{\bf U}}
\def\p{{\bf p}}

\def\P{{\bf P}}
\def\B{{\bf B}}

\def\a{{\bf a}}

\def\PPi{{\bf \Pi}}
\def\f{{\bf f}}

\rightline{DAMTP }
\rightline{M/C TH 98/24}
\centerline{\fourteenbf PERTURBATIVE EVOLUTION AND REGGE BEHAVIOUR}
\bigskip
\centerline{J R Cudell}
\centerline{Institut de Physique, Universit\'e de Li\`ege}
\vskip 5pt
\centerline{A Donnachie}
\centerline{Department of Physics, Manchester University}
\vskip 5pt
\centerline{P V Landshoff}
\centerline{DAMTP, Cambridge University$^*$}
\footnote{}{$^*$ email addresses: cudell@gw.unipc.ulg.ac.be,
\ ad@a3.ph.man.ac.uk, \ pvl@damtp.cam.ac.uk}
\bigskip
{\bf Abstract}

The known analytic properties of the Compton amplitude at small $Q^2$
place significant constraints on its behaviour at large $Q^2$. This calls 
for a re-evaluation of the role of perturbative evolution in past fits to 
data.
\vskip 15mm

{\bf 1 Introduction}

There is a widespread belief that the rise at small $x$ observed
at HERA in the proton structure function is associated with the
presence of a singularity at $N=0$ in the Mellin transform of the
DGLAP splitting function. Indeed,
fits to structure functions typically incorporate perturbative QCD 
evolution in which this singularity plays a key role\defref\dglap{
V.N. Gribov and L.N. Lipatov, Sov. J. Nucl. Phys. 15 (1972)
438 and 675\h
L.N. Lipatov, Sov. J. Nucl. Phys. 20 (1975) 94\h
Yu. L. Dokshitser, Sov. J. JETP 46 (1977) 641\h
G. Altarelli and G. Parisi, Nucl. Phys. B126 (1977) 298
},
but ignore an old\defref\short{
P V Landshoff, J C Polkinghorne and R D Short, Nuclear Physics B28 (1970) 210
}
but very basic piece of knowledge: that at small $x$ a
structure function is governed by Regge theory\defref\collins{
P D B Collins, {\it Introduction to Regge Theory and High Energy Physics},
Cambridge University Press (1977)
}. Recently, 
two of us have made\defref\twopom{
A Donnachie and P V Landshoff, Physics Letters B437 (1998) 408
}
an extremely successful fit to all small-$x$ data, from $Q^2=0$ up to
2000 GeV$^2$, in which we concentrate on the Regge theory and ignore the
requirements imposed by perturbative evolution. It is the purpose of this
paper to try to bring the two together. In particular, we argue that,
by analytically continuing in $Q^2$, one can conclude that the singularities
in the complex $N$-plane of the Mellin transform of the structure function
must already be present at small $Q^2$, and the perturbative evolution cannot
generate new singularities that appear only at high $Q^2$. In particular,
the role customarily assigned to the $N=0$ singularity is not sustainable.

Regge theory has its basis in the analyticity properties of the appropriate
scattering amplitude, in this case the virtual Compton amplitude whose
imaginary part at $t=0$ contains the familar structure functions
$F_1(x,Q^2)$ and $F_2(x,Q^2)$. A rigorous analysis of analyticity properties
of scattering amplitudes is extremely difficult\defref\sw{
R F Streater and A S Wightman, {\it PCT, Spin \& Statistics, and All That},
W A Benjamin (1964)
}. So instead one assumes\defref\elop{
R J Eden, P V Landshoff, D I Olive and J C Polkinghorne, {\it The Analytic
S-Matrix}, Cambridge University Press (1966)
} that the region of analyticity is the same 
as the analyticity region shared by all the infinite number of Feynman
graphs that one can draw for the amplitude, ignoring the fact that in a
strong-coupling r\'egime 
the {\it numerical} values of these graphs are
irrelevant because the perturbation series diverges. From this approach one may
deduce that, both for the real-photon Compton amplitude and for $Q^2<0$,
there is sufficient analyticity to derive the Regge theory --- if anything,
more so when $Q^2<0$. The analyticity properties are determined by the
masses of the {\it physical} hadrons, and it is crucial that none of these
has zero mass. The fact that gluons perhaps have zero mass is irrelevant
because they are not physical particles, even though often we pretend they
are in calculations.

With no zero-mass particle exchange, there is a finite ``Lehmann ellipse''
in the $t$-channel, within which the $t$-channel partial wave series converges
for $t>4m^2$, where $m$ is the proton mass, and for suitable values of
$s$ --- or, equivalently, $\nu$ if we are considering deep inelastic scattering.
For example, for the amplitude $T_1(\nu,t,Q^2)$ whose imaginary part at $t=0$
is $F_1(x,Q^2)$ with $x=Q^2/2\nu$,
$$
T_1(\nu,t,Q^2)=\sum _{\ell =0}^{\infty} (2\ell +1)a_{\ell}(t,Q^2) 
P_{\ell}(\cos\theta _t)
\eqno(1.1a)
$$
where $\nu$ must be such that
$$
\cos\theta _t=-{\nu-\quarter t\over \sqrt{(\quarter t+Q^2)(\quarter t-m^2)}}
\eqno(1.1b)
$$
has absolute magnitude no greater than 1.
The Froissart-Gribov technique\ref{\collins} then uses a fixed-$t$
dispersion relation to define an analytic function $a({\ell},t,Q^2)$ 
which is equal
to $a_{\ell}(t,Q^2)$ when the angular momentum $\ell$ takes the values
0,1,2,3,\dots , but is defined also for complex values of $\ell$ and 
has suitable behaviour when $|\ell |\to\infty$ to allow a Watson-Sommerfeld
transform of the partial-wave series (1.1a). This expresses the series as
an integral\footnote{$^*$}{More correctly, the Froissart-Gribov technique
splits the summation (1.1a) into two, one over even $\ell$ and the other over
odd $\ell$. Each is converted into an integral, with an ``even-signature''
amplitude $a^+({\ell},t,Q^2)$ that equals $a_{\ell}(t,Q^2)$ for 
$\ell = 0,2,4,\dots$ and an ``odd-signature amplitude $a^-({\ell},t,Q^2)$
that equals $a_{\ell}(t,Q^2)$ for $\ell = 1,3,5,\dots$.}:
$$
T_1(\nu,t,Q^2)={1\over 2i}\int_C{(2\ell +1)P_{\ell}(-\cos\theta _t)
\over\sin \pi\ell }a({\ell},t,Q^2)
\eqno(1.2)
$$
with $P_{\ell}(\cos\theta _t)$ the analytic continuation of the Legendre
polynomial to complex $\ell$. The contour $C$ is initially that of figure 1a,
wrapped arround the positive real-$\ell$ axis on which are located the zeros
of the denominator $\sin(\pi\ell )$. But the properties of $a({\ell},t,Q^2)$
allow it to be distorted to become parallel to the imaginary-$\ell$ axis,
as in figure 1b, so extending the range of values of $\nu$ and $t$ for
which the representation of $T_1(\nu,t,Q^2)$ is valid. 

\topinsert
\centerline{{\epsfxsize=125truemm\epsfbox{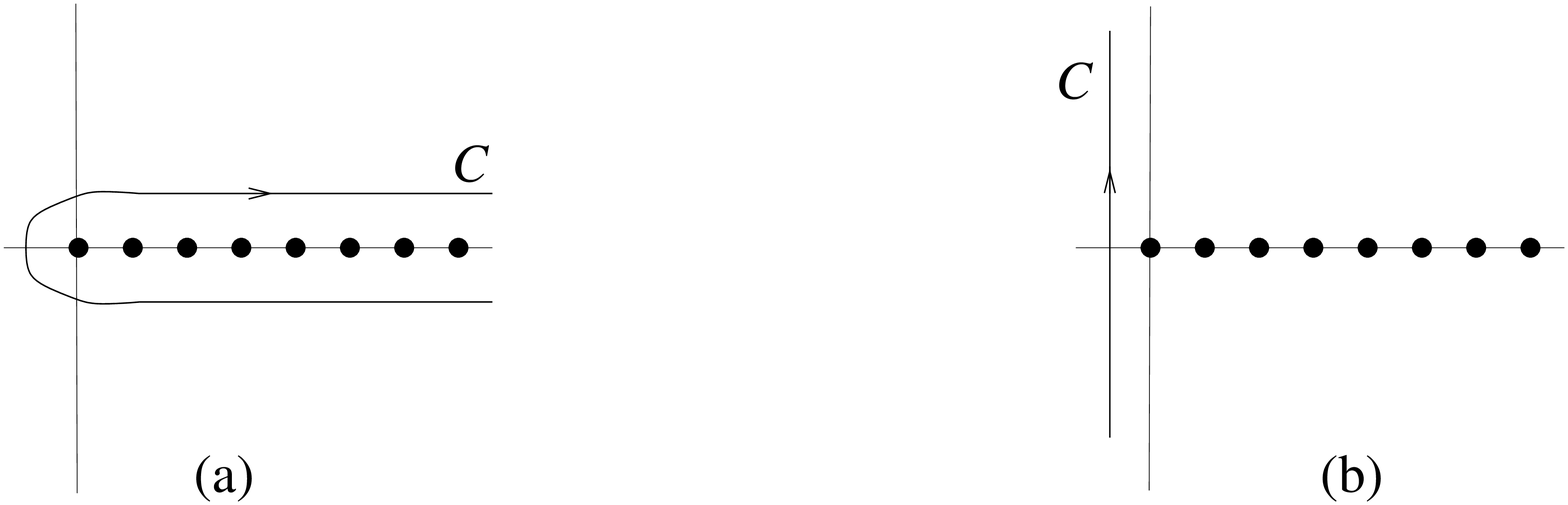}}}\hfill\break
\centerline{Figure 1: Contour $C$ in the complex $\ell$ plane for the 
integral (1.2)}
\bigskip
\centerline{{\epsfxsize=95truemm\epsfbox{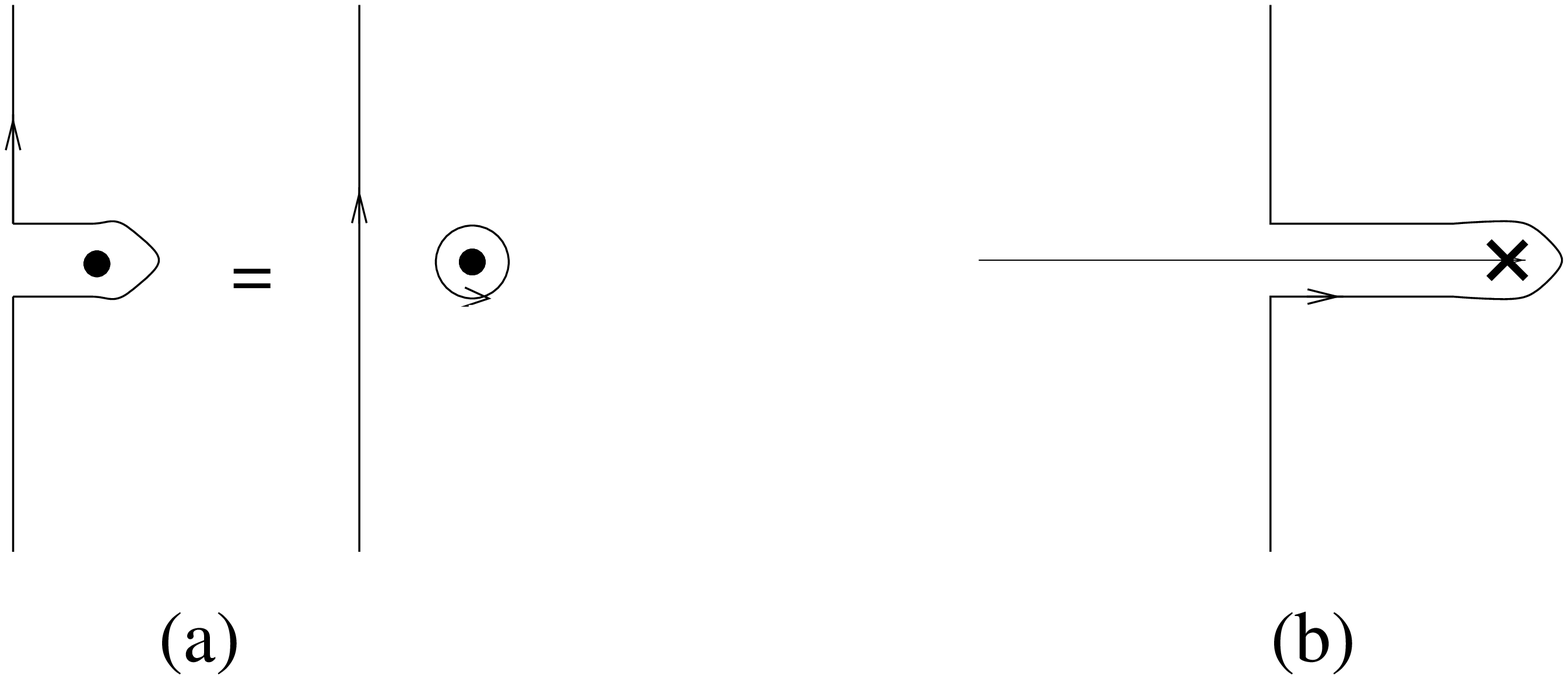}}}\hfill\break
Figure 2: Distortion of the contour $C$ caused by (a) a pole, or (b) a
branch point of $a({\ell},t,Q^2)$ crossing it.
\endinsert

{}From what is known about the analyticity structure, it is rather certain
that the positions of the singularities of $a({\ell},t,Q^2)$ in the complex
$\ell$ plane do not depend on $Q^2$, only on $t$. 
If we analytically continue in $t$ to the value we need, $t=0$, a singularity
may try to cross the contour $C$, and we must distort the contour again so
as to avoid this happening. Depending on whether the singularity is
a branch point or a pole, we then have either figure 2a or 2b. In the case
of a pole
$$
a({\ell},t,Q^2) \sim {\beta(Q^2,t)\over \ell-\alpha (t)}
\eqno(1.3)
$$
the integration around it yields a contribution 
$$
{\pi \beta(Q^2,t) P_{\alpha (t)}(\cos\theta _t)\over\sin\pi\alpha(t)}
\eqno(1.4a)
$$
to $T_1(\nu,t,Q^2)$. Now make an analytic continuation so that $\nu$ 
becomes large,
$$
\nu\gg Qm
\eqno(1.4b)
$$
Then $|\cos\theta _t|\gg 1$. For large $z$, 
$$
P_\ell(z)\sim z^{\ell}
\eqno(1.4c)
$$
so the Watson-Sommerfeld transform (1.2) becomes essentially a Mellin
transform\ref{\elop} and the ``Regge pole'' $\alpha(t)$ contributes at $t=0$
$$
b_1(Q^2)\nu^{\alpha (0)}
\eqno(1.4d)
$$
to the large-$\nu$ behaviour of $T_1(\nu,0,Q^2)$. Here, $b_1(Q^2)$ is a constant
multiple of $\beta(Q^2,0)$. In the case of
$T_2(\nu,0,Q^2)$ its definition includes a kinematic factor which reduces the
power of $\nu$ by one unit, so 
since $\nu= Q^2/x$ this gives
$$
F_2(x,Q^2)\sim f(Q^2)x^{1-\alpha (0)}
\eqno(1.5a)
$$
Regge theory gives no information about the function $f(Q^2)$, other
than that they are analytic functions with singularities whose locations
are known\ref{\elop}. The power $(1-\alpha (0))$ is independent of $Q^2$.
 
In the case where the singularity that crosses the contour $C$ is a branch
point at $\ell=\alpha _c(t)$, dragging a branch cut with it as shown in
figure 2b, the simple power of $x$ in (1.5) is replaced with
$$
F_2(x,Q^2)\sim\int ^{\alpha _c(0)}d\ell f(\ell, Q^2)x^{1-\ell}
\eqno(1.5b)
$$
which results in $F_2(x,Q^2)\sim x^{1-\alpha _c(0)}$ times an unknown function
of $Q^2$ and $\log (1/x)$. Again,
the power $(1-\alpha_c (0))$ is independent of $Q^2$.
One knows, from unitarity\ref{\collins}, that if there are poles in
the complex $\ell$ plane, there must also be branch points.
On the principle that it is usually the best strategy to try the simplest
possible assumption first, and because there is good evidence that it
is correct in purely hadronic interactions\defref\sigtot{
A Donnachie and P V Landshoff, Physics Letters B296 (1992) 227
},
in reference {\twopom} we tested the hypothesis
that the contribution to the structure function $F_2$ at small $x$
from branch points is much weaker than from poles. We made a fit of the
form
$$
F_2(x,Q^2)\sim\sum _{i=0}^2f_i(Q^2)x^{-\e _i}
\eqno(1.6a)
$$
We fixed the values of two of the powers from our knowledge\ref{\sigtot}
of hadron-hadron total cross-sections:
$$
\matrix{\e _1=0.08&\hbox{(``soft pomeron'' exchange)}\cr
\e _2=-0.45&\hbox{($\rho,\omega, f,a$ exchange)}\cr}
\eqno(1.6b)
$$
and extracted the value of  $\e _0$ from the small-$x$ data --- we used
all data for which $x<0.07$ and $Q^2$ ranged from 0 to 2000 GeV$^2$. This
gave $\e _0\approx 0.4$ and provided an excellent fit to the data over the
whole range of $Q^2$. 

\topinsert
\centerline{{\epsfxsize=85truemm\epsfbox{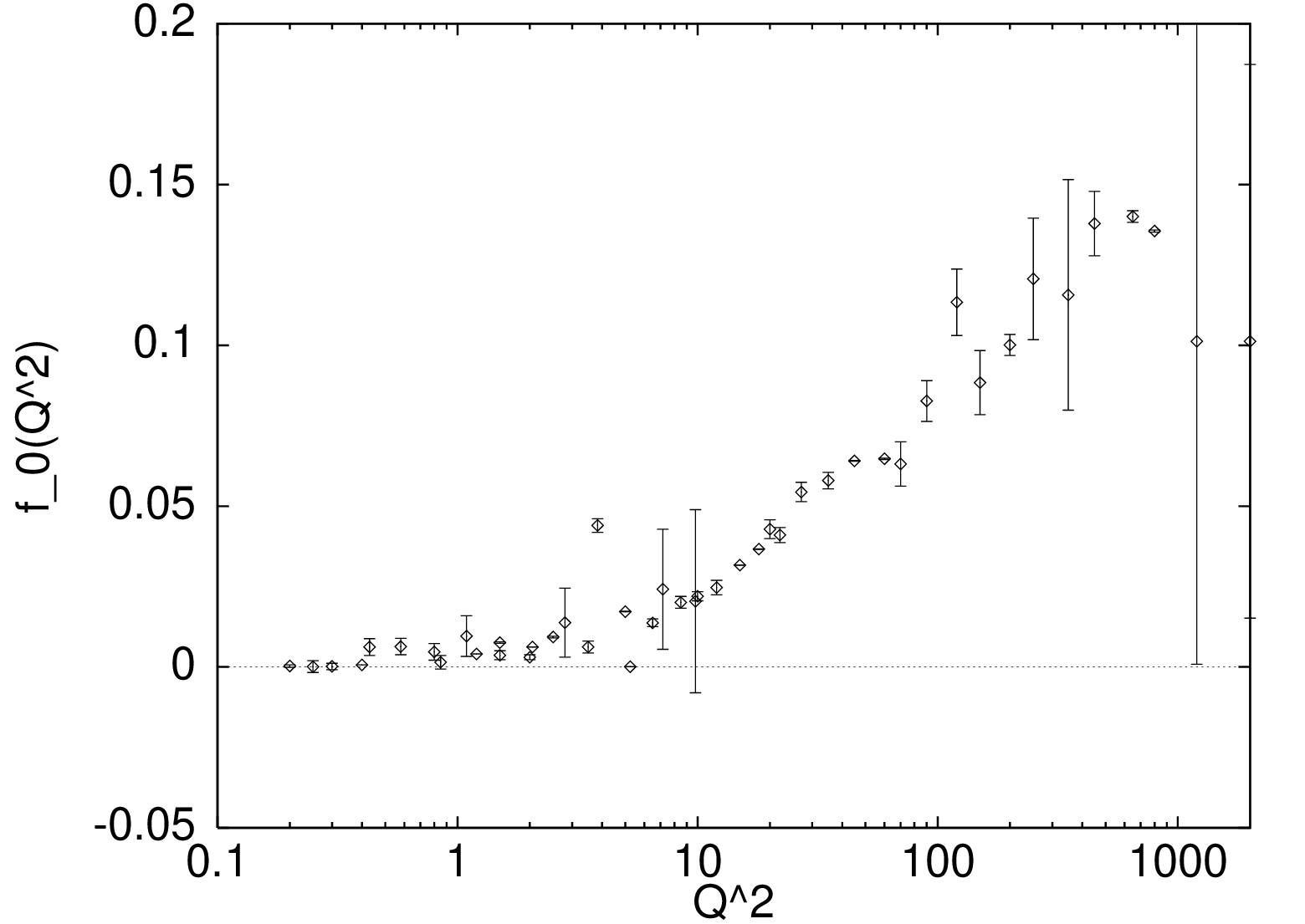}}\hfill
{\epsfxsize=85truemm\epsfbox{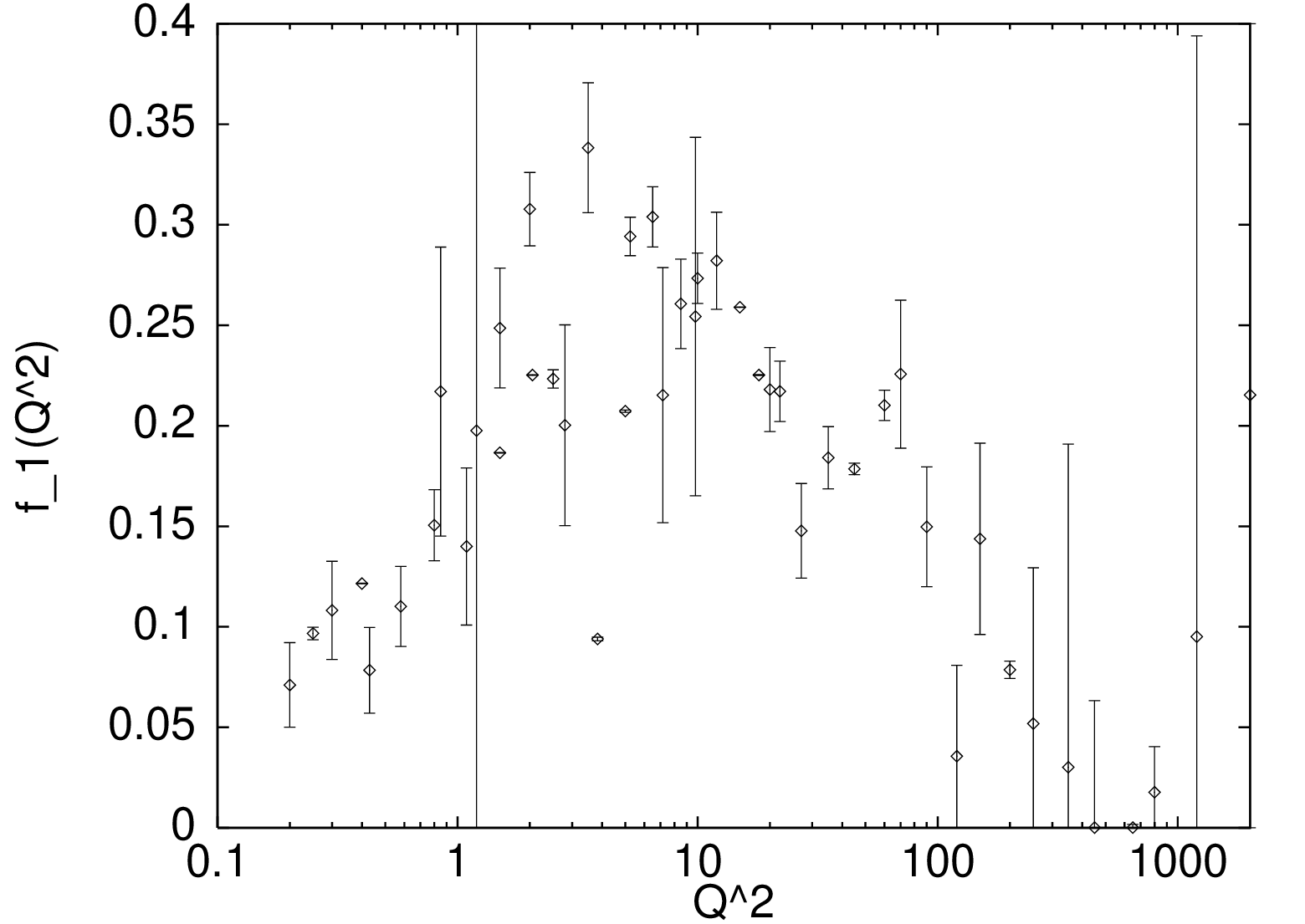}}}\hfill\break
Figure 3: The coefficient functions $f_0$ and $f_1$ extracted from
data at each $Q^2$; the error bars are from MINUIT
\endinsert

In our fit, we constrained each of the three coefficent functions
$f_i(Q^2)$ to vanish as $(Q^2)^{1+\e _i}$ as $Q^2\to 0$. Gauge invariance
requires the total $F_2$ to vanish linearly with $Q^2$ at fixed $\nu$, 
and these constraints ensure that each of the three separate 
terms in (1.6a) have this property. 
{}From the data, we found that the ``hard-pomeron'' term 
is present already
at small $Q^2$, though its  coefficient function
is very small, but it grows approximately logarithmically for
large $Q^2$. The soft-pomeron coefficient function, on the other hand,
rises quickly with $Q^2$ for small $Q^2$, peaks at a value of $Q^2$
between 5 and 10 GeV$^2$, and then becomes small at large $Q^2$.
See figure 3.
The data do not impose much constraint on how the meson-exchange coefficent
function behaves for large $Q^2$.

Regge theory gives no quantitative information about the coefficient
functions $f_i(Q^2)$, but at large $Q^2$ they should evolve according to
perturbative QCD. That is, provided they correspond to leading-twist
contributions, their behaviour at large $Q^2$ should be governed by the
DGLAP equation\ref{\dglap}. The constraint from Regge theory is that
the positions of singularities in the complex $\ell$ plane should
should {\it not} change with $Q^2$, and that new singularities do not
suddenly appear as $Q^2$ increases. In our fit to the data\ref{\twopom},
we have shown that this is fully supported by experiment.
It is our purpose in this paper to discuss it further.
\bigskip
\goodbreak
{\bf 2 The DGLAP equation: general theory}

We write the singlet DGLAP equation as\defref\esw{
R K Ellis, W J Stirling and B R Webber, {\it QCD and Collider Physics}
Cambridge University Press (1996)
}
$$
{\pd\over\pd\log Q^2}\u (x,Q^2)=\int _x^1 dz\, \bpi (z,\alpha_S(Q^2))
\,\u({x\over z},Q^2)
\eqno(2.1a)
$$
where
$$
\u= \left (\matrix{x\sum _f(q_f+\bar q_f)\cr xG\cr}\right )
\eqno(2.1b)
$$
Introduce the Mellin transform
$$
\U(N,Q^2)=\int _0^1dx\,x^{N-1}\u(x,Q^2)
\eqno(2.2a)
$$
so that
$$
\u(x,Q^2)=\int {dN\over 2\pi i}x^{-N}\U(N,Q^2)
\eqno(2.2b)
$$
As we have explained, $\U(N,Q^2)$ is essentially the complex-partial-wave
amplitude $\a ({\ell},t,Q^2)$ at $t=0$, with $\ell=N+1$, so that a power
contribution 
$$
f(Q^2)x^{-\e}
\eqno(2.3a)
$$
to $F_2(x,Q^2)$ corresponds to a pole 
$$
{\f(Q^2)\over N-\e}
\eqno(2.3b)
$$
in $\U(N,Q^2)$. If the small-$x$ behaviour of $F_2$ is (1.6a), the rightmost
singularity of $\U(N,Q^2)$ in the complex $N$-plane is a pole at $N=\e_0$,
and the integration in (2.2b) initially should be along a line
$$
\hbox{Re }N=N_0~~~~~\hbox{with }N_0>\e _0
\eqno(2.4)
$$
It is evident from its definition (2.2a) that $\U(N,Q^2)$ is analytic
to the right of this line in the complex $N$-plane, 
and for Re $N\to +\infty$ it goes to 0 for
physical values of $x$, that is $0<x<1$.
In what follows we shall analytically continue $\U(N,Q^2)$ to values
of $N$ such that Re $N<\e _0$; then it is no longer given by the representation
(2.2a), which then diverges.

Insert the Mellin-transform representation (2.2b) into the DGLAP 
equation (2.1a):
$$
\int {dN\over 2\pi i}x^{-N}{\pd\over\pd\log Q^2}\U(N,Q^2)=
\int {dN\over 2\pi i}x^{-N}\PPi (N,x,\alpha_S(Q^2)) \U(N,Q^2)
\eqno(2.5a)
$$
where
$$\PPi (N,x,\alpha_S(Q^2))=\int_x^1 dz\,z^N\bpi (z,\alpha_S(Q^2))
\eqno(2.5b)
$$
If the rightmost singularity of $\U(N,Q^2)$ in the complex $N$-plane is
a pole at $N=\e _0$, with residue $\f _0(Q^2)$, we may move the contour of
$N$ integration to the left past this pole and so pick up a contribution
to each side of (2.5a) that behaves as $x^{-\e_0}$. 
This gives the differential equation
$$
{\pd\over\pd\log Q^2}\f _0(Q^2)=\bar \PPi (\e_0,\alpha_S(Q^2)) \f _0(Q^2)
\eqno(2.6a)
$$
with $\bar \PPi (\e_0,\alpha_S(Q^2))$ the analytic continuation to
$N=\e_0$ of
$$
\bar \PPi (N,\alpha_S(Q^2))=\int_0^1 dz\,z^N\bpi (z,\alpha_S(Q^2))
\eqno(2.6b)
$$
To leading order in $\alpha_S(Q^2)$, the difference between 
$\bar\PPi (\e_0,\alpha_S(Q^2))$ and $\PPi (\e_0,x,\alpha_S(Q^2))$ behaves
as $x^{\e_0}$ at $x=0$ and therefore is negligibly small.

A differential equation such as (2.6) actually applies to the coefficient
function of each power term $\f (Q^2)x^{-\e}$ in $\u (x,Q^2)$, not just
the leading one. That is, a term that behaves as any power at some value
of $Q^2$ remains a power. However, there are apparent complications,
arising from the fact that, to lowest order in $\alpha_S(Q^2)$,
$\PPi (N,x,\alpha_S(Q^2))$ is singular at $N=0$.  We address this matter
in the next section.

In lowest order, the splitting function 
$\bpi (z,\alpha_S(Q^2))$ introduced in (2.1a) is simply proportional to
$\alpha_S(Q^2)$, and likewise its Mellin transform $\bar\PPi (N,\alpha_S(Q^2))$
of (2.6b), so that (2.6a) reads
$$
\tau{\pd\over\pd\tau}\f _0(Q^2)=\B (\e _0)\f _0(Q^2)~~~~~~~~~~~~~~~~~~~~~
\tau =\log {Q^2\over \Lambda ^2}
\eqno(2.7)
$$
The solution to (2.7) is
$$
\f _0(Q^2)=\exp \big (T(Q^2)\B (\e _0)\big )\; \f _0(Q_0^2)~~~~~~~~~~~~~~~
T(Q^2)=\log{\log Q^2/\La ^2\over \log Q_0^2/\La ^2}
\eqno(2.8)
$$
In terms of the standard anomalous-dimension matrix $\bg (j)$,
$$
\B (N)=C\left (\matrix{
\bg_{qq}(N-1) & 2N_f \bg_{qG}(N-1)\cr
\bg_{Gq}(N-1) & \bg_{GG}(N-1)\cr}\right )
~~~~~~~~~~~~~C={6\over 33-2N_f}
\eqno(2.9)
$$
The elements of $\bg (j-1)$ are plotted in figure 4.9 of reference {\esw}
whose authors have kindly provided us with the output of their computations
\footnote{$^*$}
{It is not made clear in the book that the $qg$ plot includes the
necessary factor $2N_f$.},
from which in figure 4 we plot the elements of the matrix 
$\B (N)$ in the case $N_f=4$, and in figure 5 its eigenvalues $\la _{\pm}(N)$.
\pageinsert
\centerline{{\epsfxsize=85truemm\epsfbox{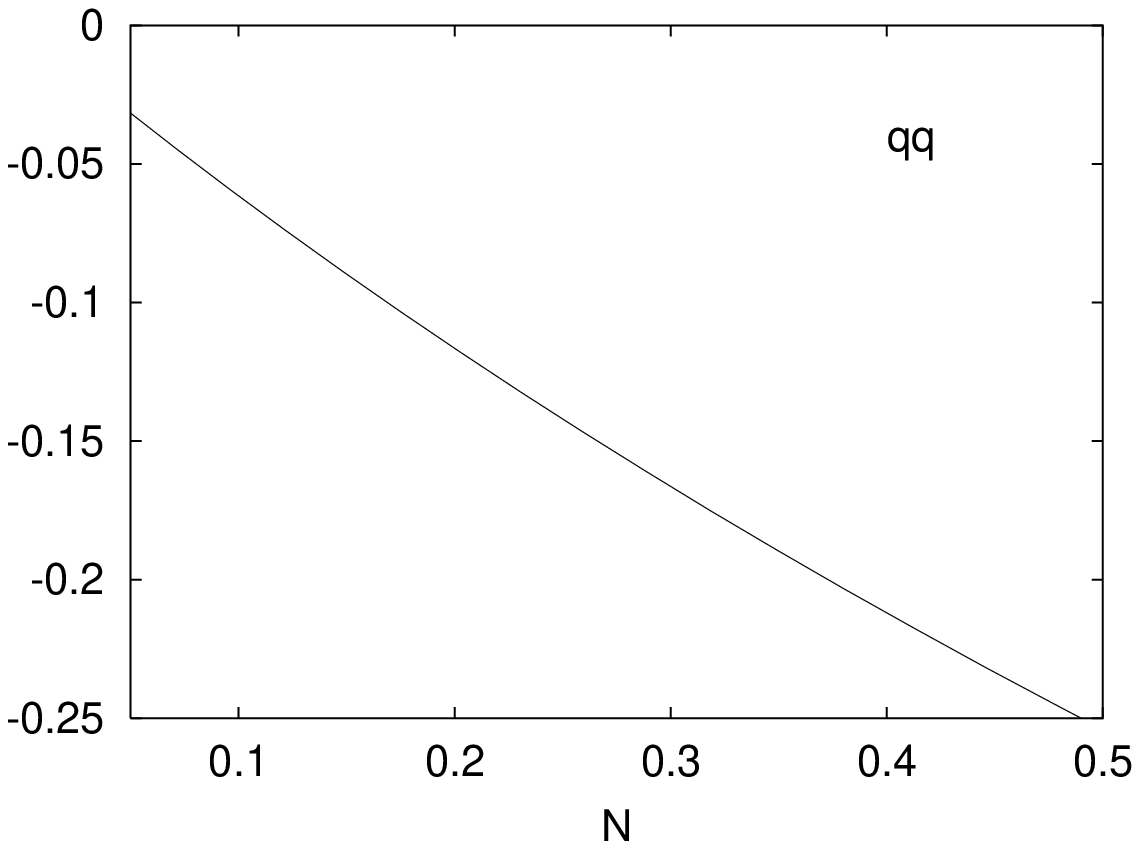}}\hfill
{\epsfxsize=85truemm\epsfbox{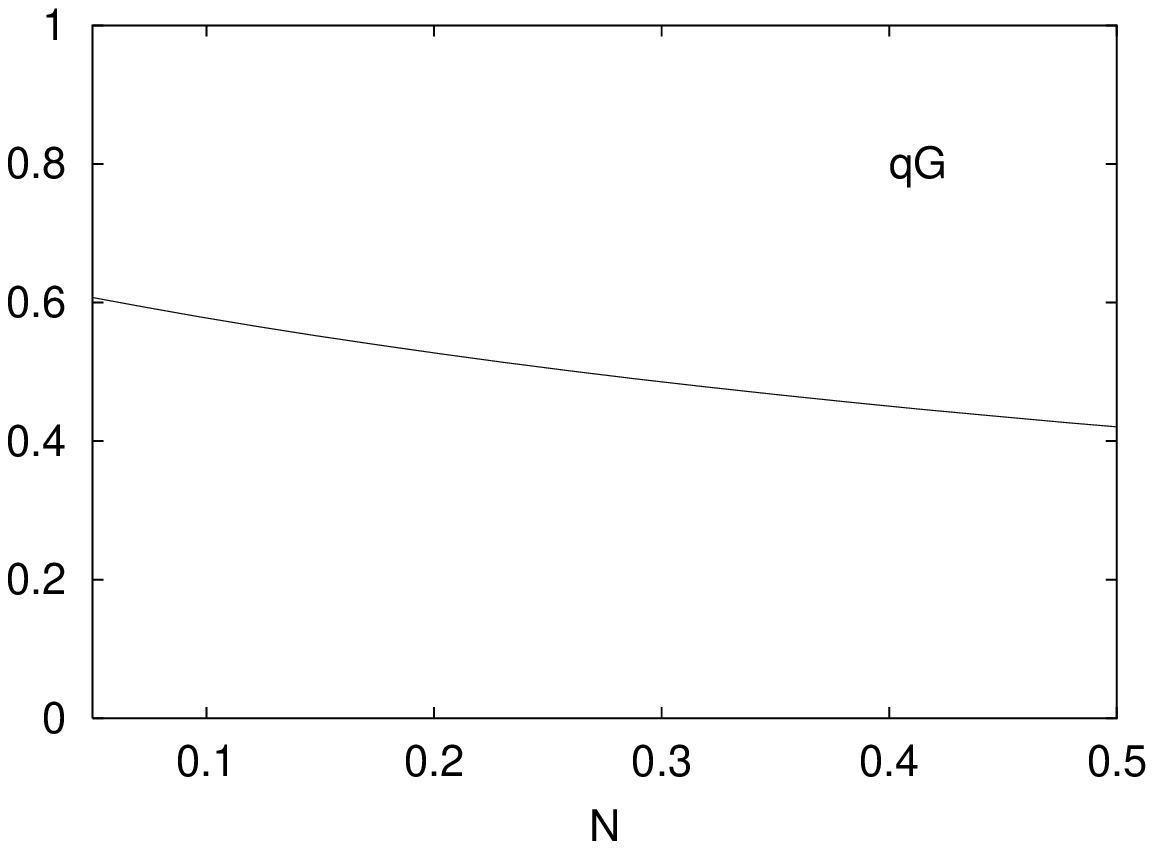}}}\hfill\break
\centerline{{\epsfxsize=85truemm\epsfbox{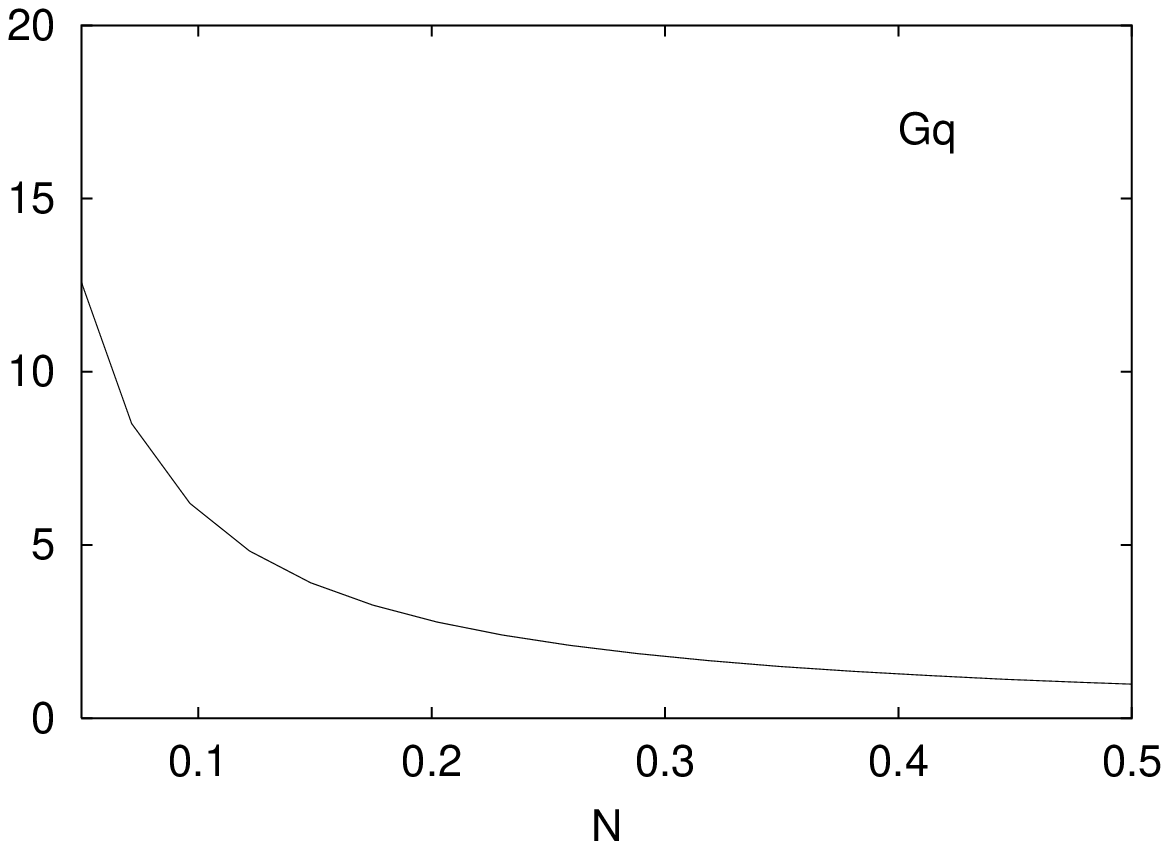}}\hfill
{\epsfxsize=85truemm\epsfbox{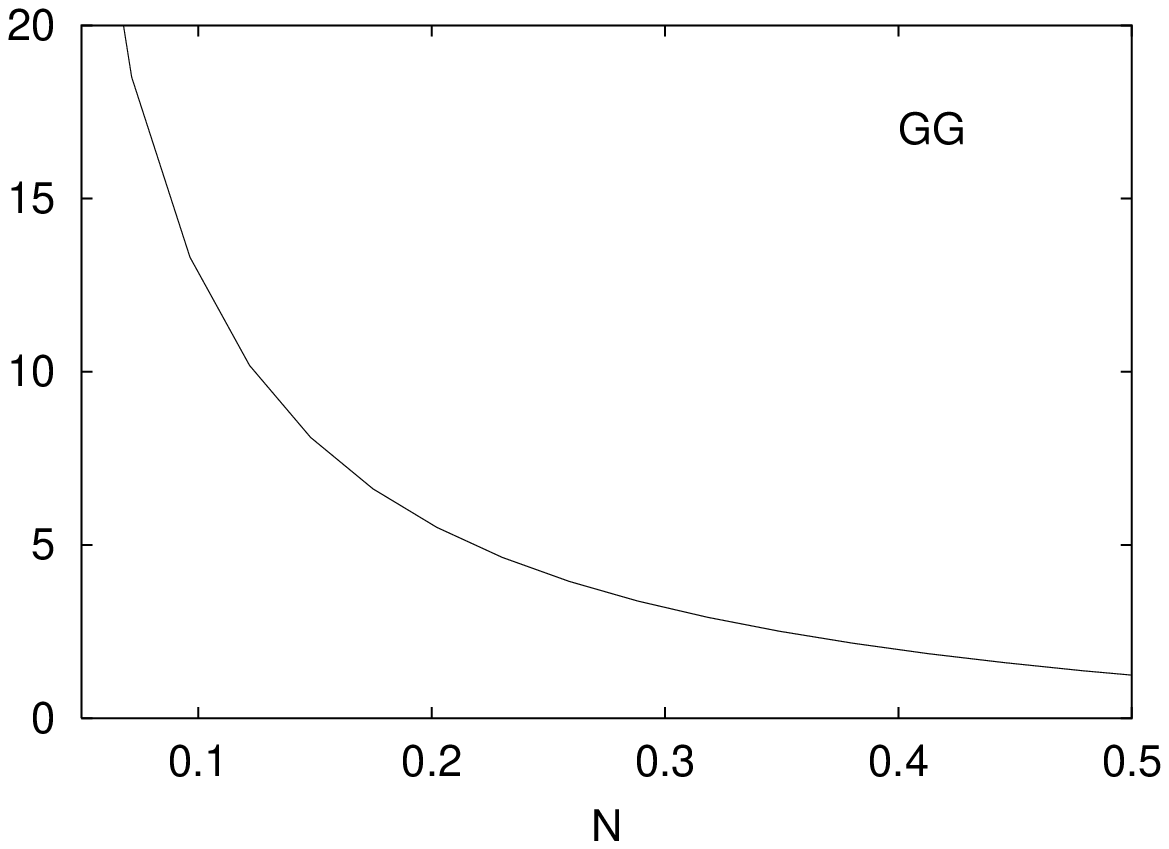}}}\hfill\break
\centerline{Figure 4: Elements of the matrix $\B (N)$}
\vskip 4mm
\centerline{{\epsfxsize=80truemm\epsfbox[65 620 270 760]{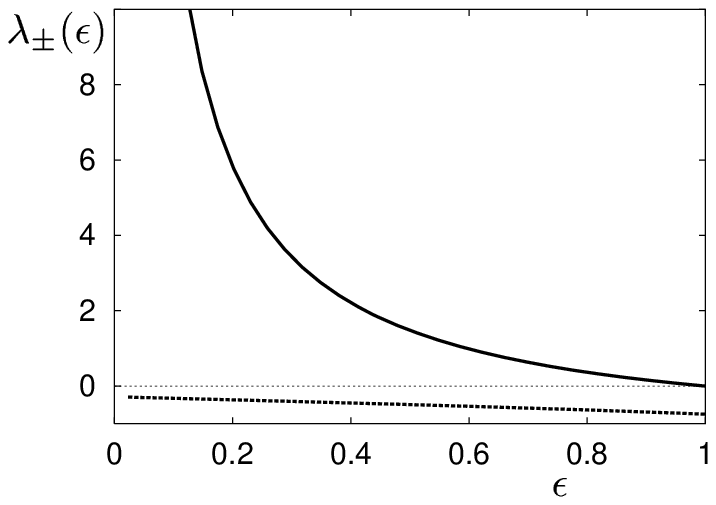}}}

\centerline{Figure 5: Eigenvalues $\la _{\pm}(N)$ of the matrix $\B (N)$}
\endinsert

If we expand $\f _0(Q^2)$ as a linear combination of the two eigenvectors
of $\B (\e _0)$:
$$
\f _0(Q^2)=\phi _{0+}(Q^2)\v _+(\e _0)+\phi _{0-}(Q^2)\v _-(\e _0)
\eqno(2.10a)
$$
where
$$
\v _{\pm}(N)=K_{\pm}(N)\left (\matrix{ B_{qG}(N)\cr \la _{\pm}(N)-B_{qq}(N)\cr}
\right )
$$$$
\big (K_{\pm}(N)\big )^{-2}=(B_{qG}(N))^2+(\la _{\pm}(N)-B_{qq}(N))^2
\eqno(2.10b)
$$
then 
$$
\phi _{0\pm}(Q^2)=
\left ({\log Q^2/\La ^2\over \log Q_0^2/\La ^2} \right )^{\la _{\pm}(\e _0)}
\phi _{0\pm}(Q_0^2)
\eqno(2.10c)
$$
According to figure 5, one eigenvalue $\la _-$ is negative, so for
large $Q^2$ only the first term on the right-hand-side of (2.10a) survives,
and then the hard-pomeron coefficient functions satisfy
$$
{f _{0q}(Q^2)\over f _{0G}(Q^2)}={B_{qG}(\e _0)\over\la _+(\e_0)-B_{qq}(\e _0)}
\eqno(2.11)
$$
when $Q^2$ is large enough. 

In our previous fit to the structure-function data we used the forms
$$
f_0(Q^2)=A_0 \left ({Q^2\over Q^2+a_0}\right )^{1+\e _0}
\left ( 1+X\log\left (1+{Q^2 \over Q_0^2}\right )\right  ) $$$$
f_1(Q^2)=A_1 \left ({Q^2\over Q^2+a_1}\right )^{1+\e _1}
 {1\over 1+\sqrt{Q^2/Q_1^2}}
$$$$
f_2(Q^2)=A_2 \left ({Q^2\over Q^2+a_2}\right )^{1+\e _2}
\eqno(2.12a)
$$
where the soft-pomeron power $\e _1$ was constrained to be 0.08 and the 
meson-exchange power $\e _2$ was fixed at -0.45. This resulted in a $\chi ^2$
just less than 1.02 per data point for the 539 data points for which $x<0.07$,
including the real-photon total cross-section measurements for which 
$\surd s >6$ GeV. The fit gave the 
``hard-pomeron''
power $\e _0$ to be about 0.4 for which, according to figure 5a, $\la _+(\e _0)$
is close to 2. On the other hand, figure 5a would give $\la _+(\e _1)$
somewhat larger, which is in obvious conflict with the behaviour of 
$f_1(Q^2)$ at large $Q^2$ extracted from the data: see figure 3b. This
reinforces our view that the soft-pomeron term is higher twist. In a new fit,
we have therefore left its form unchanged, and replaced the hard-pomeron
coefficent function with 
$$
f_0(Q^2)=A_0 \left ({Q^2\over Q^2+a_0}\right )^{1+\e _0}
\left (1+X\log\left (1+{Q^2 \over \La^2}\right )^{\la _+(\e _0)}\right  )
\eqno(2.12b)
$$
so there is one fewer parameter.
In principle, if the meson-exchange term is leading twist (which presumably
it is) we should also include a logarithmic factor in its high-$Q^2$
behaviour, raised to power $\la _+(\e _2)$ obtained by analytically
continuing $\la _+(N)$ to negative $N$. However, this term gives only a very
small contribution to the high-$Q^2$ data points and so we have left it as
in (2.12a). We find that making the change (2.12b) increases the $\chi ^2$
very slightly, to a little over 1.05 per data point.

Our fit omits any contribution from a singularity at $N=0$. We now 
explain why this is justified.
\bigskip
{\bf 3 Singularity at} $N=0$

To lowest-order in $\alpha_S$ the splitting function $\bpi (z,\alpha_S(Q^2))$
has a $z^{-1}$ singularity at $z=0$, so that its Mellin transform
$\bar \PPi (N,\alpha_S(Q^2))$, defined in (2.5b), has an $N^{-1}$
singularity at $N=0$. This means that every power $x^{-\e}$, whether
$\e$ is whether positive or negative,
generates through (2.5a) an additional singularity in $\U(N,Q^2)$
at $N=0$ as $Q^2$ increases. 
The evolution of this $N=0$ singularity is not controlled just by the
small-$x$ behaviour of the structure function. 

To investigate these matters, write the lowest-order splitting function as
$$
\bpi (z,\alpha_S(Q^2)) = 
{\alpha_S(Q^2)\over 2\pi}\big [\hat\p (z)+\p _0z^{-1}\big ]
\eqno(3.1)
$$
where $\hat\p (z)$ is regular at $z=0$.
Inserting this into (2.5) gives
$$
\int {dN\over 2\pi i}x^{-N}{\pd\over\pd\log Q^2}\U(N,Q^2)=
{\alpha_S(Q^2)\over 2\pi}\int {dN\over 2\pi i}x^{-N}\left\{\hat\P (N,x)+
N^{-1}\p _0(1-x^N)\right\} \U(N,Q^2)
\eqno(3.2)
$$
where $\hat\P (N,x)$ is a Mellin transform of $\hat\p (z)$
like (2.5b). It  vanishes so rapidly at $x=0$ that it
is a good approximation to replace it with $\hat\P (N,0)$.
Because 
$\U(N,Q^2)$ is analytic to the right of the contour 
of the $N$ integration and, from its definition (2.2a),
vanishes when $|N|\to\infty$, the last term
inside the curly bracket integrates to zero, and the inverse Mellin transform
of (3.2) is just
$$\eqalignno{
{\pd\over\pd\log Q^2}\U(N,Q^2)&\sim\alpha_S(Q^2)\left [\hat\P (N,0))+
N^{-1}\p _0\right ]\U(N,Q^2)\cr
&={\alpha_S(Q^2)\over 2\pi} \P (N)\U(N,Q^2) & (3.3)\cr}
$$
Evidently a pole (2.3b) at $N=\e$ in $\U(N,Q^2)$ gives a pole in 
$(\pd /\pd\log Q^2)\U(N,Q^2)$. 
By taking the residue at the pole of both sides of (3.3) we see that the
coefficient function $\f (Q^2)$ satisfies a differential equation
like (2.6a), involving the analytic continuation in $N$ of $\bar \P (N)$ 
to $N=\e$.  This is true for either sign of $\e$.

The solution to (3.3) is
$$
\U(N,Q^2)\sim \exp \Big [C\log{\log (Q^2/\La^2)\over \log (Q_0^2/\La^2)}
\P (N)\Big ]\U(N,Q_0^2)
\eqno(3.4)
$$
where the constant $C$ is defined in (2.9).
It is evident that, to use this equation, we must make some assumption
about $\U(N,Q_0^2)$ at some value $Q_0^2$ of $Q^2$. The common 
assumption\defref\deruj{
A De Rujula, S L Glashow, H D Politzer, S B Treiman and A Zee, 
Physical Review D10 (1974) 1649
} 
is that, for suitable choice of $Q_0^2$, it is regular at $N=0$.
Indeed, there is no reason to
suppose that, for $Q^2_0$ chosen to be 0 or very small, $\U(N,Q_0^2)$
is singular at $N=0$. That is, although the real-photon Regge amplitude
$a(\ell, t,Q^2)$ has a wrong-signature fixed pole\defref\fixed{
P V Landshoff and J C Polkinghorne, Physical Review D5 (1972) 2056
} at $\ell=1$, there is no reason to believe that at
small $Q^2$ it has also a right-signature
singularity in the spin-averaged amplitude. 
Nevertheless, as soon as $Q^2$ becomes large enough for the evolution to
be correctly described by the DGLAP equation, we learn from (3.4) that
$\U(N,Q_0^2)$ acquires an essential singularity at $N=0$. A function
such as $\U(N,Q_0^2)$ that is analytic at $N=0$ for a finite range of
$Q^2$ cannot suddenly acquire a fixed singularity at $N=0$ when it
is analytically continued in $Q^2$. 

The explanation\footnote{$^*$}{We are grateful to Guido Altarelli
and Stefano Catani for helping us to understand this}
of this contradiction is that it is wrong to
use the lowest-order approximation to the splitting function,
$\PPi (N,x,\alpha_S(Q^2))$, with its $N^{-1}$ singularity,
for any purpose where its behaviour near $N=0$ is important. The higher-order
terms in its expansion involve more and more powers of $\alpha _S /N$ and
near $N=0$ these powers become larger and larger, so that the expansion
cannot possibly converge. Our general argument, that new singularities
cannot suddenly appear in $\U (N,Q^2)$ as $Q^2$ increases, strongly
suggests that, if one were to resum the powers of $\alpha_S/N$
one would find that in fact $\PPi (N,x,\alpha_S(Q^2))$ is regular at
$N=0$. 

This is suppported by what is known about this resummation\defref\forshaw{
J Forshaw and D A Ross, {\it Quantum chromodynamics and the pomeron},
Cambridge University Press (1997)
}.
The DGLAP anomalous dimension
$\gamma _{GG}(N-1)$ in (2.9)
is equal to the BFKL anomalous dimension $\bar\gamma (N)$ calculated to
all orders in $\bar\alpha_S/N$. It is obtained by solving the equation
$$
\chi\Big ({\alpha_S\over 2\pi}\bar\gamma (N)\Big )={\pi N\over 3\alpha_S}
\eqno(3.5a)
$$
where $\chi(x)$ is the Lipatov characteristic function. To lowest order
in $\alpha_S$
$$
\chi (x)=-2\gamma_E-\psi(x)-\psi(1-x)
\eqno(3.5b)
$$
which is plotted in figure 6.
From this plot, one sees that there are two
complex-conjugate solutions for $\bar\gamma (N)$
at $N=0$; these are finite. If one expands the solution to the
implicit equations (3.5) for $\bar\gamma (N)$
in powers of $\alpha_S$,
the terms in the expansion are singular at $N=0$ because the digamma
function $\psi(\gamma)$ is singular at $\gamma =0$, but this expansion
is completely invalid near $N=0$. 

We now know\defref\fadin{
V S Fadin and L N Lipatov, Physics Letters B429 (1998) 127\h
G Camici and M Ciafaloni, Physics Letters B430 (1998) 349
}
that the lowest-order approximation (3.5b) to 
the Lipatov characteristic function is inadequate. When
the next term in the expansion is included the curve in figure 6
is pulled downwards and $\bar\gamma (N)$ is no
longer complex at $N=0$, but it is still nonsingular, as our general
argument has demanded. However, because the second term in the expansion
of the Lipatov characteristic function is so large, it is likely 
that subsequent terms are too,
and so we do not yet have a reliable estimate of just how large 
$\bar\gamma (N)$ is at $N=0$. But we know that it is nonsingular.

\topinsert
\centerline{{\epsfxsize=75truemm\epsfbox{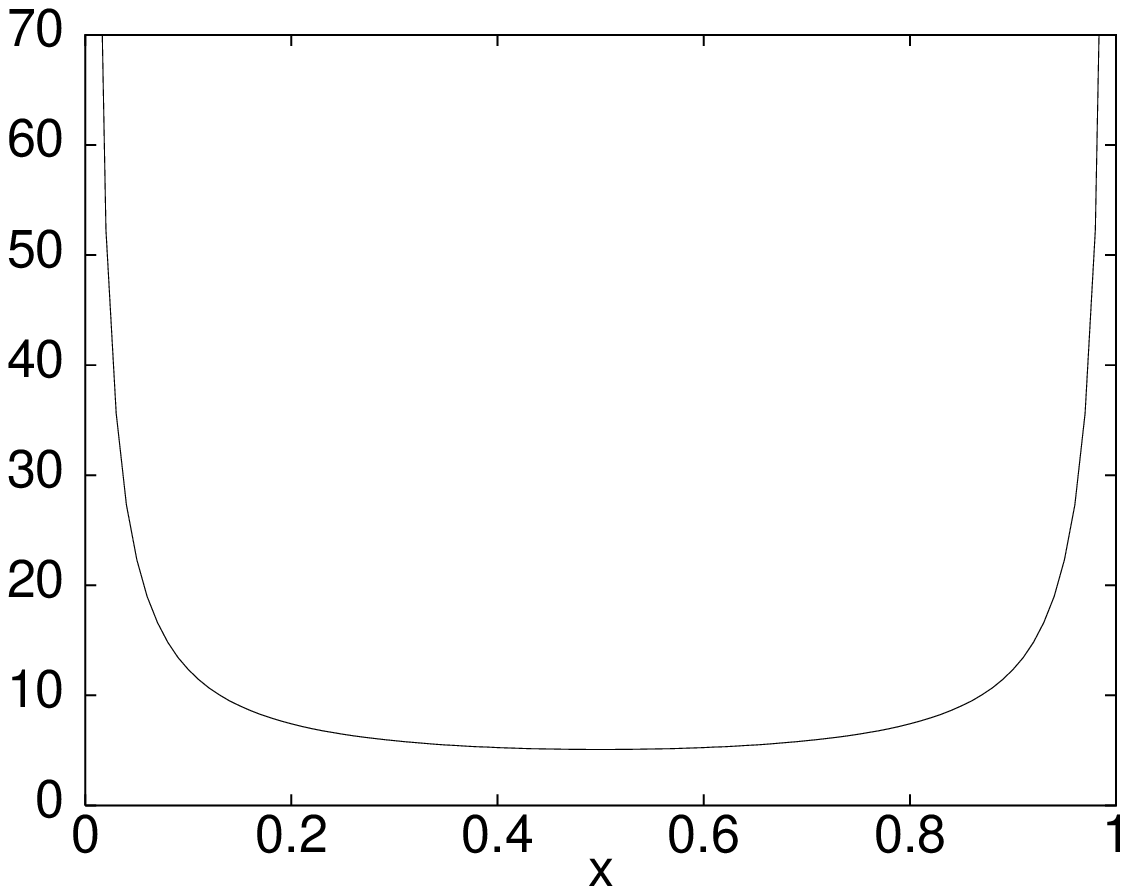}}}\hfill\break
\centerline{Figure 6: Plot of the lowest-order approximation (3.5b)
to the Lipatov characteristic function}
\endinsert

Therefore, that in fitting data one should not include 
in the splitting function any singularity at $N=0$. The only $N$-plane
singularities are those already present in the structure function
at small $Q^2$. These are the
standard singularities
of Regge theory --- the soft pomeron, the mesons, and possibly also a
hard pomeron. There may well be branch-point singularities, but 
the data are consistent with the assumption that their contribution
is relatively small at presently accessible values of $x$ and $Q^2$. 
This is the approach we followed in our previous 
paper\ref{\twopom}, and we found that it was highly successful.
We found also that the usual attitude to higher twists is more than a
little suspect: rather than making a fit that starts by assuming they are as
small as the data will allow, one should recognise that the data strongly
suggest that a significant part of the structure function at $Q^2$
values of 5~GeV$^2$ or less is higher twist --- the soft-pomeron term,
and maybe also meson exchange.
\bigskip
\goodbreak
\bigskip{\eightit
This research is supported in part by the EU Programme
``Training and Mobility of Researchers", Networks
``Hadronic Physics with High Energy Electromagnetic Probes"
(contract FMRX-CT96-0008) and
``Quantum Chromodynamics and the Deep Structure of
Elementary Particles'' (contract FMRX-CT98-0194),
and by PPARC}
\goodbreak

\medskip\immediate\closeout\rfile\writestoppt
\baselineskip=14pt{{\bf References}}\bigskip{\frenchspacing%
\parindent=20pt\escapechar=` \input refs.tmp\bigskip}\nonfrenchspacing
\bye